\documentclass[11pt]{article}
\usepackage{latexsym}
\usepackage{amsfonts}
\usepackage{epsfig}

\catcode`\@=11

\def\section{\@startsection{section}{1}{\z@}{3.5ex plus 1ex minus
 .2ex}{2.3ex plus .2ex}{\bf}}

\def\thesubsection{\arabic{section}.\arabic{subsection}}
\renewcommand{\subsection}[1]{\addtocounter{subsection}{1}
\vspace{2.5mm}\par\noindent {\it \thesubsection . #1}\par
 \vspace{0.5mm} }
\catcode`\@=12 \mathchardef\varGamma="0100
\mathchardef\varDelta="0101 \mathchardef\varTheta="0102
\mathchardef\varLambda="0103 \mathchardef\varXi="0104
\mathchardef\varPi="0105 \mathchardef\varSigma="0106
\mathchardef\varUpsilon="0107 \mathchardef\varPhi="0108
\mathchardef\varPsi="0109 \mathchardef\varOmega="010A
\def\bfone{\relax{\rm 1\kern-.35em 1}}
\DeclareFontFamily{U}{rsf}{} \DeclareFontShape{U}{rsf}{m}{n}{
  <5> <6> rsfs5 <7> <8> <9> rsfs7 <10-> rsfs10}{}
\DeclareMathAlphabet\Scr{U}{rsf}{m}{n}
\begin{document}
\begin{titlepage}
\thispagestyle{empty}
\begin{flushright}
\hfill{CERN-PH-TH/2007-015}
\end{flushright}
\vspace{35pt}
\begin{center}{ \LARGE{\bf
On the supergravity formulation of mirror symmetry in generalized Calabi-Yau manifolds}}\\
 \vspace{60pt} {\bf R. D'Auria$^\S$, S.
Ferrara$^\sharp$, M. Trigiante$^\S$}\\
\vspace{15pt}
$\S${\it Dipartimento di Fisica, Politecnico di Torino \\
C.so Duca degli Abruzzi, 24, I-10129 Torino\\
Istituto Nazionale di Fisica Nucleare, Sezione di Torino,
Italy}\\[1mm] {E-mail: riccardo.dauria@polito.it,
mario.trigiante@polito.it}\\
\vspace{15pt} $\sharp${\it CERN, Physics Department, CH 1211
Geneva 23, Switzerland.}\\ {\it INFN, Laboratori Nazionali di
Frascati, Italy}\\[1mm] {E-mail:
sergio.ferrara@cern.ch}\\
\vspace{15pt}
\begin{abstract}
We derive the complete supergravity description of the $N=2$
scalar potential which realizes a generic flux-compactification on
a Calabi-Yau manifold (generalized geometry). The effective
potential ${\Scr V}_{eff}={\Scr V}_{(\partial_Z {\Scr V}=0)}$,
obtained by integrating out the massive axionic fields of the
special quaternionic manifold, is manifestly mirror symmetric,
i.e. invariant with respect to ${\rm Sp}(2\,h_2+2)\times {\rm
Sp}(2\,h_1+2)$ and their exchange, being $h_1,\,h_2$ the complex
dimensions of the underlying special geometries. ${\Scr V}_{eff}$
has a manifestly $N=1$ form in terms of a mirror symmetric
superpotential $W$ proposed, some time ago, by Berglund and Mayr.
\end{abstract}
\end{center}
\end{titlepage}
\newpage
\baselineskip 6 mm
\section{Introduction}

Geometries which generalize Calabi-Yau manifolds in the presence
of generic fluxes \cite{gm,gmpt,h,glw1,glw2,mpt} (for
comprehensive reviews on flux compactifications see \cite{flux}),
have received considerable attention, as they realize schemes of
compactification which incorporate supersymmetry breaking and
moduli stabilization.\par On the other hand the scalar potential
originating from a compactification on such generalized geometries
can be computed, from a supergravity point of view, as a
deformation of an $N=2$ supergravity Lagrangian. This $N=2$ theory
contains hypermultiplets which define a special quaternionic
manifold $\mathcal{M}_Q$, obtained by c--map from the complex
special geometry $\mathcal{M}_{KS}$ (of dimension $h_1$)
underlying a mirror Calabi-Yau manifold \cite{mirror}. The
deformation of the $N=2$ theory is effected as an abelian gauging
of the $2h_1+3$ dimensional Heisenberg algebra of isometries of
the special quaternionic manifold \cite{heis1}. We denote by
$h_2+1$ the number of vector fields in the model, and by $h_1+1$
the number of hypermultiplets, so that $h_1=h_{11},\,h_2=h_{12}$
in Type IIB setting while $h_1=h_{12},\,h_2=h_{11}$ in Type IIA.
The resulting potential for generic fluxes
$e_I{}^\Lambda,\,e_{I\,\Lambda}$ ($I=0,\dots h_2$,
$\Lambda=0,\dots h_1$), was determined in \cite{heis2}. The
condition for an abelian gauging of the Heisenberg algebra
requires that
\begin{eqnarray}
e_{[I}{}^\Lambda\,e_{J]\,\Lambda}&=&0\,.\label{conde}
\end{eqnarray}
The generators of the Heisenberg algebra of quaternionic
isometries \cite{fs} are denoted by $X^\Lambda,\,X_\Lambda,\,{\Scr
Z}$. It is convenient to group the first $2 h_1+2$ generators in a
symplectic vector $X_A\equiv (X_\Lambda,\,X^\Lambda)$ in terms of
which the commutation relations among the Heisenberg generators
read
\begin{eqnarray}
[X_A,\,X_B]&=&2 \, {\mathbb C}_{AB}\,{\Scr Z}\,,\label{heis}
\end{eqnarray}
all the other commutators vanishing. We have denoted by
$\mathbb{C}$ the symplectic invariant matrix
\begin{eqnarray}
\mathbb{C}&=&\left(\matrix{{\bf 0} &\bfone\cr -\bfone &{\bf 0}
}\right)\,.\label{mathbbc}
\end{eqnarray}
The adjoint  action of the remaining quaternionic isometries on
the $X_A$ generators preserves this symplectic structure. These
isometries comprise those of the special K\"ahler submanifold
$\mathcal{M}_{KS}$ of the quaternionic manifold, of complex
dimension $h_1$. The generators $X_A$ are parametrized by $(2
h_1+2)$-dimensional ${\rm Sp}(2h_1+2)$-vector of axions
$Z^A=(\zeta^\Lambda,\,\tilde{\zeta}_\Lambda)$, originating from
the ten dimensional R-R forms, while the central charge  ${\Scr
Z}$ is parametrized by the axion $a$ dual to the Kalb-Ramond
antisymmetric 2-form $B_{\mu\nu}$. The electric fluxes
$e_I{}^A=(e_I{}^\Lambda,\,e_{I\,\Lambda})$, together with an
additional vector $c_I$, can be viewed as the electric components
of and embedding tensor \cite{dwst} which defines the gauge
generators $T_I$ as linear combinations of $X_A,\,{\Scr Z}$
\begin{eqnarray}
T_I&=&e_I{}^A T_A+c_I\,{\Scr Z}\,.\label{ti}
\end{eqnarray}
In what follows we shall suppose that $h_2<h_1$ and moreover that
the rectangular matrix $e_I{}^A$ have maximal rank $h_2+1$. The
gauge transformation rules for the axionic fields read
\begin{eqnarray}
\delta Z^A&=&\xi^I\,e_I{}^A\,\,\,;\,\,\,\,\delta a =
\xi^I\,c_I+\xi^I\,e_I{}^\Lambda\,\tilde{\zeta}_\Lambda-\xi^I\,e_{I\,\Lambda}\zeta^\Lambda=\xi^I\,c_I+\xi^I\,e_I{}^A\,\mathbb{C}_{AB}\,Z^B\,\,,
\end{eqnarray}
where $\xi^I(x)$ are the gauge parameters: $\delta
A^I_\mu=\partial_\mu \xi^I$. In the Type IIA framework the
 entries $e_{I}{}^A$ with $I>0$ can be characterized as geometric fluxes describing a deformation of the Calabi-Yau
 cohomology and $e_0{}^A$ as the components of the NS-NS 3-form field strength $H^{(3)}$ along the basis of 3-forms
 labelled by $A$ \cite{glw2,heis1}. The parameters $c_I$ are interpreted as R-R fluxes associated with the forms
 $F^{(0)},\,F^{(2)},\,F^{(4)},\,F^{(6)}$ in the Type IIA setting, and with the 3-form $F^{(3)}$ in the Type IIB setting.
 \par
On the other hand, in order to have a symplectic covariant
formulation of this gauging we need to dualize $h_2+1$ axions, out
of the $h_1+1$ $Z^A$, to antisymmetric tensor fields, along the
lines of \cite{tensor}. This will allow us to introduce the
magnetic counterpart $m^{IA},\,c^I$ to $e_I{}^A,\,c_I$. For an
interpretation of these parameters in terms of generalized
Calabi-Yau geometry see \cite{glw2}. An other way for introducing
magnetic fluxes would be to use the duality covariant formulation
in \cite{dwst} which describes at the same time the scalar fields
and their tensor duals, coupled to both electric and magnetic
vector fields. This procedure would eventually  require a gauge
fixing to be made and to solve certain non-dynamic equations. In
next section we shall choose a different approach consisting in
dualizing  axions parametrizing abelian quaternionic isometries
while keeping the theory covariant with respect to both the
symplectic structures on $\mathcal{M}_{SK}$ (i.e. with respect to
the group ${\rm Sp}(2h_2+2)$ of electric-magnetic duality
transformations) and on $\mathcal{M}_{KS}$ (i.e. with respect to
the group ${\rm Sp}(2h_1+2)$ acting on $Z^A$). It is convenient to
group the electric and magnetic fluxes $e_I{}^A,\,m^{IA}$ into a
single $(2 h_2+2)\times (2 h_1+2)$ rectangular flux matrix $Q$
\begin{eqnarray}
Q& \equiv &(Q_r{}^A)=\left(\matrix{e_{I}{}^A\cr
m^{IA}}\right)\,\,\,\,\,\,(r=1,\dots, 2h_2+2)\,,
\end{eqnarray}
and introduce the symplectic vector of  R-R fluxes
$c_r=(c_I,\,c^I)$. \footnote{Here we shall use the same symbol
$\mathbb{C}$ to denote the ${\rm Sp}(2h_1+2)$-invariant matrix
$\mathbb{C}_{AB}$ and the ${\rm Sp}(2h_2+2)$-invariant matrix
$\mathbb{C}_{rs}$, both having the form (\ref{mathbbc}), though
different dimensions. Which of the two matrices the symbol
$\mathbb{C}$ refers to will be clear from the context, in
particular from the dimension of the object it multiplies.} These
parameters define a $2h_2+2$ dimensional symplectic vector of
gauge generators $T_r=Q_r{}^A\,X_A+c_r\,{\Scr Z}$. The abelianity
condition $[T_r,\,T_s]=0$ now implies
\begin{eqnarray}
(Q_r{}^A
Q_s{}^B\mathbb{C}_{AB})=Q\,\mathbb{C}\,Q^T=0\,,\label{condQ0}
\end{eqnarray}
while consistency of the theory with electric and magnetic charges
requires \cite{dwst,tensor,df}
\begin{eqnarray}
(Q_r{}^A
Q_s{}^B\mathbb{C}^{rs})=Q^T\,\mathbb{C}\,Q=0\,\,;\,\,\,\,(
c_r\mathbb{C}^{rs}Q_s{}^A)=c^T\mathbb{C}Q=0\,.\label{condQ1}
\end{eqnarray}
The above conditions were found in \cite{glw2,heis2,dftup}. We
shall also use the quantity
$\tilde{Q}=\mathbb{C}^T\,Q\,\mathbb{C}=(Q^r{}_A)$.
 Let us anticipate the main result of the paper, namely the
${\rm Sp}(2\,h_2+2)\times {\rm Sp}(2\,h_1+2)$-invariant expression
of the $N=2$ scalar potential ${\Scr V}$. We shall denote by $z^a$
($a=1,\dots, h_1$) and by $w^i$  ($i=1,\dots, h_2$) the complex
scalars parametrizing $\mathcal{M}_{KS}$, submanifold of
$\mathcal{M}_Q$, and $\mathcal{M}_{SK}$ respectively. Moreover let
$V_1^A(z,\bar{z})$ and $V_2^r(w,\bar{w})$ denote the covariantly
constant symplectic sections on $\mathcal{M}_{KS}$ and
$\mathcal{M}_{SK}$ respectively. The scalar potential reads
\begin{eqnarray}
{\Scr V}&=&-\frac{1}{8\,\phi^2}\,(c+2\,Q\,\mathbb{C}\,
Z)^T\,\mathbb{C}^T\,{\Scr M}({\Scr
N}_{SK})\,\mathbb{C}\,(c+2\,Q\,\mathbb{C}\,
Z)-\nonumber\\&&-\frac{2}{\phi}\,\overline{V}_1^{\,\,T}\tilde{Q}^T{\Scr
M}({\Scr N}_{SK})\,\tilde{Q}
V_1-\frac{2}{\phi}\,\overline{{V}}_2^{\,\,T}\,Q\,{\Scr M}({\Scr
N}_{KS})\,Q^T\,V_2-\nonumber\\&&-\frac{8}{\phi}\,\overline{V_1}^{\,\,T}\,\mathbb{C}^T\,Q^T\,(V_2\,\overline{V}^{\,\,T}_2+\overline{V}_2V^T_2
)\,Q\,\mathbb{C}\,V_1\,,\label{mv0}
\end{eqnarray}
where ${\Scr M}({\Scr N})$ denotes the (negative definite)
symplectic matrix constructed in terms of the real and imaginary
part of the period matrix ${\Scr N}$ on a special K\"ahler
manifold \cite{cdf}. It then follows that the terms in the first
two lines of (\ref{mv0}) are non-negative.  Note that scalar
potential depends on $Z^A$ only through the combinations
$Q\mathbb{C} Z\equiv (Q_r{}^A\mathbb{C}_{AB}Z^B)$ which do not
contain $h_2+1 $ axions,  since it is gauge invariant, provided
the matrix $Q$ satisfies (\ref{condQ0}). These are precisely the
axions that are dualized to antisymmetric tensor fields which
acquire mass, in virtue of the  anti-Higgs mechanism, by eating
the vector fields. The combinations $Q\mathbb{C} Z$ turn out to
depend only on $h_2+1$ of the undualized axions, which then
acquire mass from the potential and can be integrated out. The
remaining $2(h_1-h_2)$ R-R scalars are flat directions. They are
absent for a self-mirror manifold, characterized by having
$h_1=h_2$. In this case $Q$ is a square matrix. The condition
which fixes the $h_2+1$ axions at the extremum value is
$c+2\,Q\,\mathbb{C}\, Z=0$. After the massive axions $Z^A$ are
integrated out we find the effective potential
\begin{eqnarray}
{\Scr V}_{eff}(\phi,w,\bar{w},z,\bar{z})&=&{\Scr
V}_{|\frac{\partial{\Scr V}}{\partial Z^A}=0}=\nonumber\\
&&-\frac{2}{\phi}\,\overline{V}_1^{\,\,T}\tilde{Q}^T{\Scr M}({\Scr
N}_{SK})\,\tilde{Q}
V_1-\frac{2}{\phi}\,\overline{{V}}_2^{\,\,T}\,Q\,{\Scr M}({\Scr
N}_{KS})\,Q^T\,V_2-\nonumber\\&&-\frac{8}{\phi}\,\overline{V_1}^{\,\,T}\,\mathbb{C}^T\,Q^T\,(V_2\,\overline{V}^{\,\,T}_2+\overline{V}_2V^T_2
)\,Q\,\mathbb{C}\,V_1\,.
\end{eqnarray}
This potential is manifestly mirror symmetric, namely symmetric if
we exchange $\mathcal{M}_{SK}$ with $\mathcal{M}_{KS}$ and replace
$Q$  by $\tilde{Q}^T$. It is now possible to show, and we shall do
it in the last section, that $V_{eff}$ has an $N=1$ form with
superpotential given by
\begin{eqnarray}
W&=&e^{-\frac{K_{SK}+K_{KS}}{2}}\,V_2(w,\bar{w})^T\,Q\,\mathbb{C}\,{V}_1(z,\bar{z})\,,
\end{eqnarray}
which coincides with the expression proposed in \cite{bm}, and
K\"ahler potential of the form
\begin{eqnarray}
K_{tot}&=&K_S+K_{SK}+K_{KS}\,,\nonumber\\
K_S&=&-\log(i(S-\bar{S}))\,\,\,;\,\,\,\,K_{SK}=-\log(i\,\overline{V}_1^{\,\,T}\mathbb{C}V_1)\,\,;\,\,\,K_{KS}=-\log(i\,\overline{V}_2^{\,\,T}\mathbb{C}V_2)\,,\label{KK}
\end{eqnarray}
$K_{SK}$ and $K_{KS}$ being the K\"ahler potentials on
$\mathcal{M}_{SK}$ and $\mathcal{M}_{KS}$ respectively.\par
 The paper is organized as follows. In section \ref{s2} we perform the
dualization of the axion $a$ and of those components of $Z^A$
which transform non trivially under the gauge group. We then
introduce the magnetic components of the embedding tensor in the
resulting Lagrangian. In section \ref{s3} we extend the results of
\cite{heis2}, using the general formulae of \cite{tensor,dwst}, to
write the full ${\rm Sp}(2\,h_2+2)\times {\rm
Sp}(2\,h_1+2)$-invariant scalar potential. Finally in section
\ref{s4} we make contact with the $N=1$ potential proposed in
\cite{bm}. We end with some conclusions.

\section{Dualization with electric and magnetic charges}
\label{s2} Let us start by introducing the notations. We consider
a special quaternionic manifold $\mathcal{M}_Q$ of real dimension
$4\,(h_1+1)$, which is parametrized by the scalars
\begin{eqnarray}
q^u&=&\{\phi,\,a,\,\zeta^\Lambda,\,\tilde{\zeta}_\Lambda,\,z^a\}\,,
\end{eqnarray}
where, from Type IIB point of view, $a$ is the scalar dual to the
2--form NS tensor $B_{\mu\nu}$, $\zeta^0=C_{(0)}$,
$\zeta^\Lambda=C_{(2)}^\Lambda$, ($\Lambda>0$), $\tilde{\zeta}_0$
is dual to $C_{\mu\nu}$, $\tilde{\zeta}_\Lambda=C_{(4)\,\Lambda}$,
($\Lambda>0$), $\phi$ describes the four--dimensional dilaton and
the complex scalars $z^a$ are the K\"ahler moduli of the
Calabi-Yau and span the special K\"ahler submanifold
$\mathcal{M}_{KS}$ of complex dimension $h_1$. In the Type IIA
description the axions $\zeta^\Lambda,\,\tilde{\zeta}_\Lambda$
arise as the components of the R-R 3-form along a basis
$\alpha_\Lambda,\,\beta^\Lambda$ of the third chomology group
$H^{(3)}$ of the Calabi-Yau, while $z^a$ describe its complex
structure moduli. We can introduce on $\mathcal{M}_{KS}$ the
projective coordinates $\mathcal{X}^\Lambda$ which define the
upper components of a holomorphic symplectic section:
$\mathcal{X}^0=1,\,\mathcal{X}^a=z^a$. As anticipated in the
introduction, there exists a subgroup of the isometry group
generated by a Heisenberg algebra $(X_A,\,{\Scr
Z})\equiv(X_\Lambda,\,X^\Lambda,{\Scr Z} )$, whose action of the
hyperscalars has the following form:
\begin{eqnarray}
\delta\zeta^\Lambda&=&\alpha^\Lambda\,,\nonumber\\
\delta\tilde{\zeta}_\Lambda&=& \beta_\Lambda\,,\nonumber\\
\delta a &=&
\gamma+\alpha^\Lambda\tilde{\zeta}_\Lambda-\beta_\Lambda\zeta^\Lambda\,,
\label{transf}\end{eqnarray} and which close the algebra
(\ref{heis}). Using the notations of \cite{fs}, we introduce the
following one forms
\begin{eqnarray}
v&=&e^{\tilde{K}}\,[d\phi-i\,(da+\tilde{\zeta}^T\,
d\zeta-\zeta^T\, d\tilde{\zeta})]\,,\nonumber\\
u&=&2i\,e^{\frac{\tilde{K}+\hat{K}}{2}}\,
\mathcal{X}^T\,(d\tilde{\zeta}-{\Scr N}_{KS}\,d\zeta)\,, \nonumber\\
E&=&i\,e^{\frac{\tilde{K}-\hat{K}}{2}}\,P\,N^{-1}\,(d\tilde{\zeta}-{\Scr N}_{KS}\,d\zeta)\,, \nonumber\\
e&=&P\,d\mathcal{X}\,,\label{uvforms}
\end{eqnarray}
where
\begin{eqnarray}
e^{\tilde{K}}&=&\frac{1}{2\phi}=\frac{e^{2\,\varphi}}{2},\,\,\,;\,\,\,\,\,e^{\hat{K}}=
\frac{1}{2\overline{\mathcal{X}}^T
N\mathcal{X}}=\frac{e^{K_{KS}}}{2}\,\,\,;\,\,\,\,\,\,(\phi>0)\,,
\end{eqnarray}
where $\varphi$ denotes the four dimensional dilaton and $K_{KS}$
is the K\"ahler potential on $\mathcal{M}_{KS}$ defined in
(\ref{KK}).
\par
The metric on the quaternionic manifold reads:
\begin{eqnarray}
ds^2&=& \bar{v}\,v+\bar{u}\,u+\bar{E}\,E+\bar{e}\,e=\nonumber\\&&
K_{a\bar{b}}\,d
z^a\,d\bar{z}^{\bar{b}}+\frac{1}{4\,\phi^2}\,(d\phi)^2+
\frac{1}{4\,\phi^2}\,(d a+ d Z^T\mathbb{C}
Z)^2-\frac{1}{2\,\phi}\,d Z^T\,{\Scr M}({\Scr N}_{KS})\,d
Z\,,\label{ul}
\end{eqnarray}
where ${\Scr N}_{KS}$ is the period matrix on $\mathcal{M}_{KS}$
\footnote{In our conventions ${\Scr N}_{KS}=i\,{\Scr N}_s$ where
${\Scr N}_s$ is the period matrix used in \cite{fs}.}, the
symplectic matrix ${\Scr M}({\Scr N})$ is defined as
follows:\begin{eqnarray} {\Scr M}({\Scr N})&=&\left(\matrix{\bfone
& -{\rm Re} {\Scr N}\cr 0 & \bfone }\right)\left(\matrix{{\rm Im}
{\Scr N}& 0\cr 0 & {\rm Im} {\Scr N}^{-1}
}\right)\left(\matrix{\bfone &0\cr -{\rm Re} {\Scr N}&
\bfone}\right)\,,
\end{eqnarray}
and the axion vector $Z^A=\left(\matrix{\zeta^\Lambda\cr
\tilde{\zeta}_\Lambda}\right)$ was defined in the
introduction.\par The Killing vectors associated with the abelian
gauge algebra generators $T_I$ defined in (\ref{ti}) read:
\begin{eqnarray}
k_I&=&(c_I+e_I{}^\Lambda\,\tilde{\zeta}_\Lambda-e_{I\Lambda}\,\zeta^\Lambda)\,\frac{\partial}{\partial
a }+e_I{}^\Lambda\,\frac{\partial}{\partial \zeta^\Lambda
}+e_{I\Lambda}\,\frac{\partial}{\partial \tilde{\zeta}_\Lambda
}\,.
\end{eqnarray}
 Let us start with the deformation \cite{heis1} of the quaternionic Lagrangian (\ref{ul}) which corresponds to the
 chosen gauging of the Heisenberg isometry algebra:
\begin{eqnarray}
{\Scr L}&=&-K_{a\bar{b}}\,
dz^a\wedge\star\,d\bar{z}^{\bar{b}}-\frac{1}{4\,\phi^2}\,(Da-Z^A\,\mathbb{C}_{AB}\,DZ^B)\wedge
\star
(Da-Z^A\,\mathbb{C}_{AB}\,DZ^B)+\nonumber\\&&+\frac{1}{2\,\phi}\,DZ^A\,{\Scr
M}({\Scr N}_{KS})_{AB}\wedge\star
DZ^B\,,\nonumber\\&&\label{ludua}
\end{eqnarray}
where the covariant derivatives are defined as follows:
\begin{eqnarray}
Da&=&da-c_I\,A^I-e_I{}^A\,\mathbb{C}_{AB}\,Z^B\,A^I\,,\nonumber\\
DZ^A&=&dZ^A-e_I{}^A\,A^I\,,
\end{eqnarray}
The electric charges $e_I{}^A$ satisfy the  cocycle condition
(\ref{conde}) corresponding to the requirement that the gauge
algebra be abelian:
\begin{eqnarray}
e_I{}^A\,e_J{}^B\,\mathbb{C}_{AB}&=&0\,.\label{cobo}
\end{eqnarray}
As a consequence of the above condition the charges $e_I{}^A$
select an abelian ``section'' of the Heisenberg algebra to be
gauged. Using $e_I{}^A$, we can split the RR scalar fields in two
orthogonal sets $Z^I,\,\hat{Z}^A$, as follows:
\begin{eqnarray}
Z^A&=&e_I{}^A\,Z^I+\hat{Z}^A\,.
\end{eqnarray}
It is also useful to define the scalars $Z_I\equiv
e_I{}^A\mathbb{C}_{AB}Z^B=e_I{}^A\mathbb{C}_{AB}\hat{Z}^B$. We may
define the above splitting in a more formal way by introducing a
matrix $\tilde{e}_A{}^I$ satisfying the conditions
\begin{eqnarray}
\tilde{e}_A{}^I\,e_I{}^B&=&P^{(+)}{}_A{}^B\,\,;\,\,\,\tilde{e}_A{}^I\,e_J{}^A=\delta_I{}^J\,,
\end{eqnarray}
where $P^{(+)}{}_A{}^B$ is the projector on the $h_2+1$
dimensional subspace corresponding to the non vanishing minor of
$e_I{}^A$. We also define the orthogonal projector
$P^{(-)}{}_A{}^B=\delta_A^B-P^{(+)}{}_A{}^B$. Using these
projectors we can define
$Z^I=\tilde{e}_{A}{}^I\,P^{(+)}{}_B{}^A\,Z^B$ and
$\hat{Z}^A=P^{(-)}{}_B{}^A\,Z^B$. Note that under gauge
transformations
\begin{eqnarray}
\delta Z^I&=&\xi^I\,\,;\,\,\,\delta\hat{Z}^A=0\,,
\end{eqnarray}
namely the $\hat{Z}^A$ components are gauge invariant. In other
words the embedding tensor $e_I{}^A,\,c_I$ defines an abelian
subalgebra of the Heisenberg algebra spanned by the axions
$a,\,Z^I$. Our  aim is to dualize these scalars. We start from
rewriting the vielbein along the ${\Scr Z}$ direction on the
tangent space, in the following form
\begin{eqnarray}
d a+ d Z^T\mathbb{C} Z&=& da
+Z_I\,dZ^I-Z^I\,dZ_I-\hat{Z}^A\mathbb{C}_{AB}\,d\hat{Z}^B\,.\label{va}
\end{eqnarray}
From the above expression we see that, if we make the redefinition
$a\rightarrow a+Z_I\,Z^I$, all the scalars $Z^I$ in eq.
(\ref{va}), and therefore also in (\ref{ludua}),  can be covered
by derivatives and thus $a$ and $Z^I$ can be dualized into closed
3--forms $H=dB,\,H_I=dB_I$. To this end we introduce a set of
unconstrained 1--forms $\eta, U^I$ replacing the differentials
$da, dZ^I$ in the Lagrangian (\ref{ludua}) and add the 3--forms
$H,\,H_I$ as Lagrange multipliers. Note that the $H_I$ can be
expressed as combinations of $2\,(h_1+1)$ 3-forms $H_A$ and
similarly the corresponding antisymmetric tensors $B_I$ can be
expressed as combinations of $2\,(h_1+1)$ 2-forms $B_A$:
\begin{eqnarray}
H_I&=&e_I{}^A\,H_A\,\,\,;\,\,\,\,B_I=e_I{}^A\,B_A\,\,\,;\,\,\,\,H_A=dB_A\,.
\end{eqnarray}
The resulting
 first order Lagrangian reads:
\begin{eqnarray}
{\Scr L}_Q&=&-K_{a\bar{b}}\,
dz^a\wedge\star\,d\bar{z}^{\bar{b}}-\frac{1}{4\,\phi^2}\,(\eta+2\,Z_I\,U^I-R)\wedge
\star\, (\eta+2\,Z_I\,U^I-R)+\nonumber\\&&
+(U^I-A^I)\,\Delta_{IJ}\,\wedge \star (U^J-A^J)+2\,
(U^I-A^I)\,e_I{}^A\,\Delta_{AB}\,\wedge \star\,
d\hat{Z}^B+d\hat{Z}^A\,\Delta_{AB}\,\wedge
\star\, d\hat{Z}^B+\nonumber\\
&& +H\wedge (\eta-da)+H_I\wedge (U^I-dZ^I)\,,\label{ldua}
\end{eqnarray}
where we have used the following notation:
\begin{eqnarray}
R&=&2\,Z_I\, A^I+c_I\,A^I+\hat{Z}^A\,\mathbb{C}_{AB}\, d\hat{Z}^B\,,\nonumber\\
\Delta_{AB}&=&\frac{1}{2\,\phi}\,{\Scr M}({\Scr
N}_{KS})_{AB}\,\,;\,\,\,\,\Delta_{IJ}=e_I{}^A
e_J{}^B\,\Delta_{AB}\,.
\end{eqnarray}
By varying the Lagrangian with respect to $a$ and $Z^I$ we obtain
$H=dB,\,H_I=dB_I$. The field equations from the variations with
respect to $U^I$ and $\eta$ are:
\begin{eqnarray}
\frac{\delta{\Scr L}}{\delta \eta}=0&\Rightarrow
&\eta+2\,Z_I\,U^I-R=-2\phi^2\star\,H\,,\nonumber\\
\frac{\delta{\Scr L}}{\delta U^I}=0&\Rightarrow
&Z_I\,(\eta+2\,Z_J\,U^J-R)=2\,\Delta_{IJ}\,\phi^2\,(U^J-A^J)+2\,\phi^2\,e_I{}^A\,\Delta_{AB}\,d\hat{Z}^B-
\nonumber\\&&-\phi^2\,\star\,H_I\,.
\end{eqnarray}
Solving the above equations with respect to $\eta,\, U_I$ and
substituting in the first order Lagrangian we obtain the dual
Lagrangian:
\begin{eqnarray}
{\Scr L}_{QD}&=&-K_{a\bar{b}}\,
dz^a\wedge\star\,d\bar{z}^{\bar{b}}-(\phi^2-\Delta^{IJ}\,Z_I\,Z_J)\,
H\wedge\, \star\,
H+\frac{1}{4}\,\Delta^{IJ}\,H_I\wedge\star\,H_J-\Delta^{IJ}\,H\wedge\star
H_I\,Z_J-\nonumber\\&&-(H_I-2\,H\,Z_I)\,\Delta^{IJ}\,e_J{}^A\,\Delta_{AB}\wedge
d\hat{Z}^B+H\wedge
\hat{Z}^A\,\mathbb{C}_{AB}\,d\hat{Z}^B+(H_I+c_I\,H)\,
\wedge\,A^I+\nonumber\\&&+
d\hat{Z}^A\,\tilde{\Delta}_{AB}\,\wedge\star\,d\hat{Z}^B\,,
\end{eqnarray}
where
\begin{eqnarray}
\Delta^{IK}\,\Delta_{KJ}&=&\delta^I_J\,\,;\,\,\,\tilde{\Delta}_{AB}=\Delta_{AB}-\Delta^{IJ}\,e_J{}^C\,\Delta_{CA}\,
\,e_I{}^D\,\Delta_{DB}
\end{eqnarray}
The dual Lagrangian is invariant under the following gauge
transformations:
\begin{eqnarray}
\delta A^I&=&d\xi^I\,\,\,;\,\,\,\,\,\delta
B_I=d\Xi_I\,\,;\,\,\,\delta B=d\Xi\,,
\end{eqnarray}
where the 1-forms $\Xi_I,\,\Xi$ parametrize the tensor-gauge
transformations. We can complete the Lagrangian (\ref{ldua}) by
adding the kinetic and theta term of the vector fields:
\begin{eqnarray}
{\Scr L}_{vec}&=&{\rm Im}({\Scr N}_{SK})_{IJ}\, F^I\wedge\star\,
F^J+\frac{1}{2}\,{\rm Re}({\Scr N}_{SK})_{IJ}\,F^I\wedge\, F^J\,.
\end{eqnarray}
It is straightforward to generalize the above construction by
including magnetic charges $m^{IA},\,c^I$, according to the
following prescription \cite{tensor}:
\begin{itemize}
\item{In ${\Scr L}_{vec}$ substitute $F^I$ by $\hat{F}^I\equiv F^I+m^{IA}\,B_A+c^I\,B$.}
\item{ In ${\Scr L}_{QD}$ substitute the topological term $H_I\wedge\,
A^I=e_I{}^A\,H_A\wedge\,A^I =-e_I{}^A\,B_A\wedge F^I$ by
$-e_I{}^B\,B_B\wedge(\hat{F}^I-\frac{1}{2}\,m^{IA}\,B_A-\frac{1}{2}\,c^I\,B)$.
The same for the term $-c_I\,B\wedge F^I$.}
\end{itemize}
In conclusion the final Lagrangian describing scalar, tensor and
vector fields coupled to each other by means of electric and
magnetic charges reads:
\begin{eqnarray}
{\Scr L}_D&=&{\rm Im}({\Scr N}_{SK})_{IJ}\, \hat{F}^I\wedge\star\,
\hat{F}^J+\frac{1}{2}\,{\rm Re}({\Scr
N}_{SK})_{IJ}\,\hat{F}^I\wedge\,
\hat{F}^J-\nonumber\\&&-K_{a\bar{b}}\,
dz^a\wedge\star\,d\bar{z}^{\bar{b}}-(\phi^2-\Delta^{IJ}\,Z_I\,Z_J)\,
H\wedge\, \star\,
H+\frac{1}{4}\,\Delta^{IJ}\,H_I\wedge\star\,H_J-\Delta^{IJ}\,H\wedge\star
H_I\,Z_J-\nonumber\\&&-(H_I-2\,H\,Z_I)\,\Delta^{IJ}\,e_J{}^A\,\Delta_{AB}\wedge
d\hat{Z}^B+H\wedge
\hat{Z}^A\,\mathbb{C}_{AB}\,d\hat{Z}^B-\nonumber\\&&-(B_I+c_I\,B)\,
\wedge\,(\hat{F}^I-\frac{1}{2}\,m^{IA}\,B_A-\frac{1}{2}\,c^I\,B)+
d\hat{Z}^A\,\tilde{\Delta}_{AB}\,\wedge\star\,d\hat{Z}^B\,.\label{lag}
\end{eqnarray}
The above Lagrangian enjoys the extra tensor--gauge invariance:
\begin{eqnarray}
\delta B_I=d\Xi_I\,\,;\,\,\,\,\delta B=d\Xi\,\,\,;\,\,\,\,\delta
A^I=-m^{IA}\, \Xi_A-c^I\,\Xi\,,
\end{eqnarray}
provided the following conditions are met:
\begin{eqnarray}
e_I{}^A\,m^{IB}-e_I{}^B\,m^{IA}&=&0\,\,;\,\,\,\,c_I\,m^{IB}-e_I{}^B\,c^I=0\,,\label{em}
\end{eqnarray}
which are equivalent to (\ref{condQ1}). The form of Lagrangian
(\ref{lag}) is consistent with the construction given in
\cite{tensor}\footnote{In \cite{tensor} to role of the indices
$I,\,\Lambda$ is exchanged.} as far as the kinetic metric of the
tensors and the tensor--scalar couplings are concerned. This is
the case since, although we introduce $2\,h_1+2$ tensors $B_A$
formally corresponding to all of the symplectic scalars $Z^A$,
only the combination $B_I=e_I{}^A{}B_A$ and $B$ are actually
propagating and they mirror the scalars $Z^I,\, a$ which
parametrize an abelian subalgebra of the Heisenberg algebra, due
to condition (\ref{cobo}). A related observation is the fact that
in paper \cite{tensor} the choice of dualizing the parameters of
an abelian algebra was made from the very beginning so that
condition (\ref{cobo}) was not needed. Let us note that also the
combination $m^{IA}\,B_A$ can be expressed in terms of the only
propagating tensors $B_I$. Indeed we can write
\begin{eqnarray}
m^{IA}\,B_A&=&m^{JA}\,e_J{}^B\,\tilde{e}_B{}^I\,B_A=m^{JB}\,e_J{}^A\,\tilde{e}_B{}^I\,B_A=m^{JB}\,\tilde{e}_B{}^I\,B_J\,,
\end{eqnarray}
where the first of conditions (\ref{em}) has been used.
\section{Scalar potential with electric and magnetic fluxes}
\label{s3} The general form of the ${\Scr N}=2$ scalar potential
is \cite{n2}:
\begin{equation}
{\Scr V} = 4\,  h_{uv} k^u_I k^v_J \,L^I\,\overline{L}^J+
g_{r\bar{s}}\, k^r_I k^{\bar{s}}_J \,L^I\, \overline{L}^J+
(U^{IJ}-3\,L^I \overline{L}^J){\Scr P}^x_I\,{\Scr P}^x_J\,,
\end{equation}
where the second term does not contribute to the gauging we are
considering, which involves quaternionic isometries only since it
is abelian. The vectors $L^I$ denote the upper part of the
covariantly holomorphic symplectic section $V$ on the special
K\"ahler manifold $\mathcal{M}_{SK}$ parametrized by the vector
multiplet scalars $w^i,\,\bar{w}^{\bar{\imath}}$. The expression
for the momentum maps ${\Scr P}^x_J$ is:
\begin{eqnarray}
{\Scr P}^x_I&=&k_I^u\,\omega^x_u\,,\label{p}
\end{eqnarray}
where $\omega^x$ is the ${\rm SU}(2)$ connection. This form is
Heisenberg--invariant and so is therefore the ${\rm SU}(2)$
curvature. This justifies the absence of a compensator on the
right hand side of eq. (\ref{p}).\par
 It is useful to rewrite the
scalar potential in two equivalent ways:
\begin{eqnarray}
{\Scr V}&=&4\,  h_{uv}\, k^u_I\, k^v_J \,L^I\,\overline{L}^J+
(U^{IJ}-3\,L^I \overline{L}^J)\,k^u_I\,
k^v_J\,\omega^x_u\,\omega^x_v\,,\label{V1}\\
{\Scr V}&=&-\frac{1}{2}\,({\rm Im}{\Scr N}_{SK})^{-1 IJ}\,k^u_I
k^v_J\,\omega^x_u\,\omega^x_v + 4\,( h_{uv}-
\omega^x_u\,\omega^x_v )\,k^u_I \,k^v_J
\,L^I\,\overline{L}^J\,,\label{V2}
\end{eqnarray}
where we have used the special geometry identity:
\begin{eqnarray}
U^{IJ}&=&-\frac{1}{2}\,({\rm Im}{\Scr N}_{SK})^{-1
IJ}-\overline{L}\,L^T\,.
\end{eqnarray}
  In
order to evaluate the expression on the right hand side of eq.
(\ref{V2}) it is useful to compute the following quantity
\cite{fs}:
\begin{eqnarray}
G_{IJ}&=&k_I^u\,k_J^v\,(h_{uv}-\omega^x_u\,\omega^x_v)=k_I^u\,k_J^v\,
[\bar{v}\,v+\bar{u}\,u+\bar{E}\,E-(\bar{v}\,v+4\,\bar{u}\,u)]_{uv}\,.
\end{eqnarray}
Using the following notation:
\begin{eqnarray}
r_I&=&c_I+2\,(e_{I}{}^{\Lambda}\,\tilde{\zeta}_\Lambda-
e_{I\Lambda}\,\zeta^\Lambda)\,\,\,;\,\,\,\,s_{I\Lambda}=e_{I\Lambda}-e_I{}^\Sigma\,
({\Scr N}_{KS})_{\Sigma\Lambda}\,,
\end{eqnarray}
we can express $G_{IJ}$ as follows:
\begin{eqnarray}
G_{IJ}&=& 2 \,
e^{\tilde{K}}\,\bar{s}_{I\Lambda}\,s_{J\Sigma}\,\left({\Scr
U}-3\,\overline{{\mathcal L}}\,{\mathcal L}^T
\right)^{\Lambda\Sigma}\,\,;\,\,\,\,{\Scr U}=-\frac{1}{2}\,({\rm
Im}{\Scr N}_{KS})^{-1}-\overline{{\mathcal L}}\,{\mathcal
L}^T\,\,\,;\,\,\, {\mathcal
L}=e^{\frac{K_{KS}}{2}}\,\mathcal{X}\,.
\end{eqnarray}
In deriving the above expression for $G_{IJ}$ we made use of the
following properties:
\begin{eqnarray}
N^{-1}P^\dagger P N^{-1}&=&e^{K}\,(-N^{-1}+{\mathcal
L}\,\overline{{\mathcal L}}^T)\,,\nonumber\\
-\frac{1}{2}\,({\rm Im}{\Scr N}_{KS})^{-1}&=&-N^{-1}+{\mathcal
L}\,\overline{{\mathcal L}}^T+\overline{{\mathcal L}}\,{\mathcal
L}^T\,.
\end{eqnarray}
Now we can evaluate the two equivalent expressions for the scalar
potential given in eqs. (\ref{V1}) and (\ref{V2}) \cite{heis2}:
\begin{eqnarray}
{\Scr
V}&=&\overline{L}^I\,{L}^J\,\left[\frac{1}{\phi^2}\,(c_I+2\,e_I\mathbb{C}
Z)\,(c_J+2\,e_J\mathbb{C} Z)-\frac{2}{\phi}\,e_I\,{\Scr M}({\Scr
N}_{KS}
)\,e_J^T\right]+\nonumber\\&&\frac{1}{2\,\phi}\,(U-3\,\bar{L}
{L}^T)^{(IJ)}\,\left(\frac{1}{2\,\phi}\,r_I\,r_J+8\,
\bar{s}_{I\Lambda}\,s_{J\Sigma}\,\overline{{\mathcal L}}^\Lambda
{\mathcal L}^\Sigma \right)\,,\label{V12}\\
{\Scr V}&=&-\frac{1}{4\,\phi}\,({\rm Im}{\Scr N}_{SK})^{-1
IJ}\,\left(\frac{1}{2\,\phi}\,r_I\,r_J+8\,
\bar{s}_{I\Lambda}\,s_{J\Sigma}\,\overline{{\mathcal L}}^\Lambda
{\mathcal L}^\Sigma \right)+\nonumber\\
&&\frac{4}{\phi}\,\overline{L}^I\,{L}^J\,\bar{s}_{(I|\Lambda}\,s_{J)\Sigma}\,\left({\Scr
U}-3\,\bar{{\mathcal L}}\,{\mathcal L}^T
\right)^{\Lambda\Sigma}\,,\label{V22}
\end{eqnarray}
where we have introduced the following vectors:
$e_I=\left(\matrix{e_I{}^\Lambda\cr e_{I\Lambda}}\right)$. The
first equation (\ref{V12}) is useful for those gaugings which
involve just the graviphoton $A^0_\mu$, e.g. Type IIA with NS flux
or Type IIB on a half--flat ``mirror'' manifold \cite{gm}. Indeed
in these cases the term in the second line of (\ref{V12}) does not
contribute for cubic special geometries in the vector multiplet
sector since:
\begin{eqnarray}
(U-3\,\overline{L} {L}^T)^{00}&=&0\,.
\end{eqnarray}
Similarly the expression (\ref{V22}) is of particular use for
those gaugings which involve only isometries $\Lambda=0$, like for
instance Type IIA on a half--flat manifold or Type IIB on the
``mirror'' manifold with NS flux since, for cubic special
quaternionic geometries:
\begin{eqnarray}
\left({\Scr U}-3\,\overline{{\mathcal L}}\,{\mathcal L}^T
\right)^{00}&=&0\,\,\Rightarrow\,\,\,e^{K_{KS}}=-\frac{1}{8}\,
({\rm Im}{\Scr N}_{KS})^{-1\,00}.
\end{eqnarray}
Let us now rewrite the scalar potential ${\Scr V}$ as a symplectic
covariant form in terms of the electric and magnetic charge matrix
$Q\equiv (Q_r{}^A)$ defined in the introduction. To this end we
use the covariantly holomorphic symplectic sections $V_2$ and
$V_1$,  associated with $\mathcal{M}_{SK}$  and $\mathcal{M}_{KS}$
respectively:
\begin{eqnarray}
V_2&=&(V^r_2)=\left(\matrix{L^I\cr
M_I}\right)\,\,\,;\,\,\,\,V_1=(V_1^A)=\left(\matrix{\mathcal{L}^\Lambda\cr
\mathcal{M}_\Lambda}\right)\,.
\end{eqnarray}
Using the properties
\begin{eqnarray}
\bar{s}_{I\Lambda}\,({\rm Im}{\Scr
N}_{KS})^{-1\Lambda\Sigma}\,s_{I\Sigma}&=&e_I{}^A\,{\Scr M}({\Scr N}_{KS})_{AB}\,e_I{}^B\,,\nonumber\\
{s}_{I\Lambda}\,\mathcal{L}^\Lambda
&=&-e_{I}{}^A\,\mathbb{C}_{AB}\,V_1^B\,,
\end{eqnarray}
the scalar potential ${\Scr V}$ in (\ref{V12}), or equivalently in
(\ref{V22}), has the following ${\rm Sp}(2\,h_2+2)$ invariant
extension
\begin{eqnarray}
{\Scr V}&=&-\frac{1}{8\,\phi^2}\,(c+2\,Q\,\mathbb{C}\,
Z)^T\,\mathbb{C}^T\,{\Scr M}({\Scr
N}_{SK})\,\mathbb{C}\,(c+2\,Q\,\mathbb{C}\,
Z)-\nonumber\\&&-\frac{2}{\phi}\,\overline{V}_1^{\,\,T}\tilde{Q}^T{\Scr
M}({\Scr N}_{SK})\,\tilde{Q}
V_1-\frac{2}{\phi}\,\overline{{V}}_2^{\,\,T}\,Q\,{\Scr M}({\Scr
N}_{KS})\,Q^T\,V_2-\nonumber\\&&-\frac{8}{\phi}\,\overline{V_1}^{\,\,T}\,\mathbb{C}^T\,Q^T\,(V_2\,\overline{V}^{\,\,T}_2+\overline{V}_2V^T_2
)\,Q\,\mathbb{C}\,V_1\,,\label{mv}
\end{eqnarray}
where $c$ denotes the symplectic vector of R-R electric and
magnetic charges defined in the introduction:
$c\equiv(c_I,\,c^I)$. Note that ${\Scr V}$ depends only on the
gauge invariant component $\hat{Z}^A$ of $Z^A$ and not on the
$Z^I$ which have been dualized to tensor fields, in virtue of the
property (\ref{condQ0})
\begin{eqnarray}
Q_r{}^A\,\mathbb{C}_{AB}\,Z^B&=&Q_r{}^A\,\mathbb{C}_{AB}\,e_I{}^B\,Z^I+Q_r{}^A\,\mathbb{C}_{AB}\hat{Z}^A=Q_r{}^A\,\mathbb{C}_{AB}\hat{Z}^A\,.
\end{eqnarray}
The equation of motion for $\hat{Z}$ imply the following condition
\begin{eqnarray}
c+2\,Q\,\mathbb{C}\, \hat{Z}&=&0\,,\label{fixz}
\end{eqnarray}
which fixes part of the undualized $\hat{Z}$ axions. To illustrate
which of these axions are fixed and which are flat directions let
us choose a basis for $Z^A$ so that, if we split the upper index
$\Lambda$ in  $\Lambda=(I,{\lambda})$: ${\rm det}(e_I{}^J)\neq
0,\,e_I{}^\lambda=e_{I\,\Lambda}=0$. Conditions $Q\mathbb{C}
Q^T=Q^T\mathbb{C} Q=0$ then imply that the only non vanishing
components of $m^{I\,A}$ are described by the non singular matrix
$m^{IJ}$ satisfying the condition $m^{I[J}\,e_{I}{}^{K]}=0$. The
combinations $Q\mathbb{C} \hat{Z}$ then single out the only
scalars $\tilde{\zeta}_I$, which therefore are the only components
of the vector $Z^A$ entering the potential, and thus fixed by
condition (\ref{fixz}). Therefore in this case the fate of the
original $Z^A$ scalars is summarized as follows
\begin{eqnarray}
(h_2+1)\,\,\,Z^I\equiv \zeta^I &\longrightarrow & \mbox{dualized
to tensor
fields}\,\,\,\,B_{\mu\nu I}\,,\nonumber\\
(h_2+1)\,\,\,Z_I\equiv \tilde{\zeta}_I &\longrightarrow & \mbox{fixed by (\ref{fixz})}\,,\nonumber\\
2\,(h_1-h_2)\,\,\,\tilde{\zeta}_\lambda,\,\zeta^\lambda
&\longrightarrow & \mbox{flat directions for ${\Scr V}$}\,.
\end{eqnarray}
Upon implementation of conditions (\ref{fixz}),
 the first term in the scalar potential (\ref{mv})
 vanishes, and the resulting effective potential ${\Scr V}_{eff}$, as a function of the remaining scalar
 fields, acquires the following mirror symmetric expression
 \begin{eqnarray}
{\Scr V}_{eff}(\phi,w,\bar{w},z,\bar{z})&=&{\Scr
V}_{|\frac{\partial{\Scr V}}{\partial
Z^A}=0}=-\frac{2}{\phi}\,\overline{V}_1^{\,\,T}\tilde{Q}^T{\Scr
M}({\Scr N}_{SK})\,\tilde{Q}
V_1-\frac{2}{\phi}\,\overline{{V}}_2^{\,\,T}\,Q\,{\Scr M}({\Scr
N}_{KS})\,Q^T\,V_2-\nonumber\\&&-\frac{8}{\phi}\,\overline{V_1}^{\,\,T}\,\mathbb{C}^T\,Q^T\,(V_2\,\overline{V}^{\,\,T}_2+\overline{V}_2V^T_2
)\,Q\,\mathbb{C}\,V_1\,.\label{mv1}
 \end{eqnarray}
The above formula for ${\Scr V}$ is manifestly invariant  if we
 exchange $\mathcal{M}_{SK}$ with $\mathcal{M}_{KS}$ and
 $Q$ with $\tilde{Q}^T$.
\section{Formulation in terms of an $N=1$ superpotential}
\label{s4} In this section we show that the expression for ${\Scr
V}$ in (\ref{mv}) can be described in terms of the $N=1$
superpotential proposed in \cite{bm}
\begin{eqnarray}
W&=&e^{-\frac{K_{SK}+K_{KS}}{2}}\,V_2^T\,Q\,\mathbb{C}\,V_1\,,
\end{eqnarray}
where $K_{SK}(w,\bar{w})$ and $K_{KS}(z,\bar{z})$ are the K\"ahler
potentials on $\mathcal{M}_{SK}$ and $\mathcal{M}_{KS}$ defined in
(\ref{KK}) . The scalars of the $N=1$ theory are
$S,\bar{S},\,w^i,\bar{w}^{\bar{\imath}},\,z^a,\,\bar{z}^{\bar{a}}$
and span a K\"ahler manifold with K\"ahler potential given in
(\ref{KK}). The $N=1$ scalar potential reads
\begin{eqnarray}
{\Scr V}_{N=1}&=&e^{K_{tot}}\left(g^{a\bar{b}}\,D_a W\,D_{\bar{b}}
\overline{W}+g^{i\bar{\jmath}}\,D_i W\,D_{\bar{\jmath}}
\overline{W}+g^{S\bar{S}}\,D_S W\,D_{\bar{S}}
\overline{W}-3\,|W|^2\right)\,,
\end{eqnarray}
where the covariant derivatives are defined as $D_x W=\partial_x
W+\partial_x K_{tot}\,W$, where $x=i,a,S$. Note that $W$ is
$S$-independent and therefore
\begin{eqnarray}
g^{S\bar{S}}\,D_S W\,D_{\bar{S}} \overline{W}&=&g^{S\bar{S}}\,D_S
K_{S}\,D_{\bar{S}}K_S\,|W|^2=|W|^2\,.
\end{eqnarray}
 Let us now use the
following properties of special geometry
\begin{eqnarray}
g^{a\bar{b}}\,D_a
V_1D_{\bar{b}}\overline{V_1}&=&-\frac{1}{2}\,\mathbb{C}^T{\Scr
M}({\Scr N}_{KS})\mathbb{C}-\overline{V_1}\,V_1^T\,,\nonumber\\
g^{i\bar{\jmath}}\,D_i
{V}_2D_{\bar{\jmath}}\overline{{V}}_2&=&-\frac{1}{2}\,\mathbb{C}^T{\Scr
M}({\Scr N}_{SK})\mathbb{C}-\overline{{V}}_2\,{V}_2^T\,,
\end{eqnarray}
and write the relevant terms in $V_{N=1}$
\begin{eqnarray}
g^{a\bar{b}}\,D_a W\,D_{\bar{b}}
\overline{W}&=&e^{-\frac{K_{SK}+K_{KS}}{2}}\,\left(-\frac{1}{2}\,V_2^T\,Q\,{\Scr
M}({\Scr N}_{KS})\,Q^T\,\overline{V}_2-V_1^T\mathbb{C}^TQ^T\overline{V}_2V^T_2 Q\mathbb{C}\overline{V}_1\right)\,,\nonumber\\
g^{i\bar{\jmath}}\,D_i W\,D_{\bar{\jmath}}
\overline{W}&=&e^{-\frac{K_{SK}+K_{KS}}{2}}\,\left(-\frac{1}{2}\,V_1^T\tilde{Q}^T\,{\Scr
M}({\Scr
N}_{SK})\,\tilde{Q}\,\overline{V}_1-V_1^T\mathbb{C}^TQ^T\overline{V}_2V^T_2 Q\mathbb{C}\overline{V}_1\right)\,,\nonumber\\
-2\,|W|^2&=&-2\,e^{-\frac{K_{SK}+K_{KS}}{2}}\,V_1^T\mathbb{C}^TQ^TV_2\overline{V}^T_2
Q\mathbb{C}\overline{V}_1\,.
\end{eqnarray}
The scalar potential therefore can be recast in the following form
\begin{eqnarray}
{\Scr V}_{N=1}&=&
e^{K_S}\,\left(-\frac{1}{2}\,\overline{V}_1^{\,\,T}\tilde{Q}^T{\Scr
M}({\Scr N}_{SK})\,\tilde{Q}
V_1-\frac{1}{2}\,\overline{{V}}_2^{\,\,T}\,Q\,{\Scr M}({\Scr
N}_{KS})\,Q^T\,V_2-\right.\nonumber\\&&\left.-2\,\overline{V_1}^{\,\,T}\,\mathbb{C}^T\,Q^T\,(V_2\,\overline{V}^{\,\,T}_2+\overline{V}_2V^T_2
)\,Q\,\mathbb{C}\,V_1\right)\,,
\end{eqnarray}
which coincides with the expression in (\ref{mv}) provided ${\rm
Im} S=-\exp(-K_S)/2=-\phi/8$.
\section{Conclusions}
We have derived the scalar potential for an $N=2$ supergravity
theory with general electric and magnetic gauging of an abelian
subalgebra of the Heisenberg isometry algebra of a special
quaternionic K\"ahler manifold. Although we have only discussed
 the bosonic action, by applying the results of \cite{tensor}, the
 full Lagrangian, including fermionic terms and the transformation
 laws are known.This Lagrangian is supposed to describe the
 effective theory for a compactification of Type II superstring on
 a generalized Calabi-Yau manifold, which, in this context, is
 viewed as a deformation of a Calabi-Yau manifold when general
 fluxes are turned on. One limitation of this description is that
 classical c-map has been used to obtain a manifest ${\rm Sp}(2\,h_2+2)\times {\rm
Sp}(2\,h_1+2)$-symmetric description. It would be interesting to
describe a situation in which a quantum c-map \cite{hfsv},
encompassing both perturbative and non-perturbative effects for
the quaternionic geometry, is used in this context of generalized
geometries.

\section{Acknowledgements}
 Work supported in part by the European
Community's Human Potential Program under contract
MRTN-CT-2004-005104 `Constituents, fundamental forces and
symmetries of the universe', in which R. D'A. and M.T.  are
associated to Torino University. The work of S.F. has been
supported in part by European Community's Human Potential Program
under contract MRTN-CT-2004-005104 `Constituents, fundamental
forces and symmetries of the universe' and the contract MRTN-
CT-2004-503369 `The quest for unification: Theory Confronts
Experiments', in association with INFN Frascati National
Laboratories and by D.O.E. grant DE-FG03-91ER40662, Task C.

 \end{document}